# Shock vaporization/devolatilization of evaporitic minerals, halite and gypsum, in an open system investigated by a two-stage light gas gun

Kosuke Kurosawa[1]*, Ryota Moriwaki[1], Goro Komatsu[2,1], Takaya Okamoto[3], Hiroshi Sakuma[4], Hikaru Yabuta[5] and Takafumi Matsui[1]

[1] Planetary Exploration Research Center, Chiba Institute of Technology, 2-17-1, Tsudanuma, Narashino, Chiba 275-0016, Japan

[2] International Research School of Planetary Sciences, Università d'Annunzio, Viale Pindaro, 42, 65127 Pescara, Italy

[3] Institute of Space and Astronautical Science, Japan Aerospace Exploration Agency, 3-1-1, Yoshinodai, Chuo-ku, Sagamihara, Kanagawa 252-5210, Japan

[4] Research Center for Functional Materials, National Institute for Materials Science, 1-1, Namiki, Tsukuba, Ibaraki 305-0044, Japan

[5] Department of Earth and Planetary Systems Science, Graduate School of Science, Hiroshima University, 1-3-1, Kagamiyama, Higashi–Hiroshima, Hiroshima 739-8526, Japan

* Corresponding author: Kosuke Kurosawa (Kosuke.kurosawa@perc.it-chiba.ac.jp)

## Key Points

- Gas release from halite and gypsum was studied using a two-stage light gas gun setup that avoids contamination from gun-related gases.
- Vaporization of halite at 31 GPa and devolatilization (water loss) from gypsum at 11 GPa were detected.
- Impacts might have resulted in efficient production of perchlorates in Martian soil and produced the anhydrite found in Gale Crater.

## Abstract

Dry lakebeds might constitute large volatile reservoirs on Mars. Hypervelocity impacts onto ancient dry lakebeds would have affected the volatile distribution on Mars. We developed a new experimental method to investigate the response of evaporitic minerals (halite and gypsum) to impact shocks in an open system. This technique does not result in chemical contamination from the operation of the gas gun. The technique is termed the "two-valve method" and the gun system is located in the Planetary Exploration Research Center, Chiba Institute of Technology, Japan. We detected the vaporization of halite at 31 GPa and devolatilization from gypsum at 11 GPa, suggesting that impact-induced volatile release from dry lakebeds has periodically occurred throughout Martian history. The vaporization of halite deposits might have enhanced the production of perchlorates, which are found globally on Mars. The water loss from gypsum possibly explains the coexisting types of Ca-sulfates found in Gale Crater. **(149/150 words)**

**Plain Language Summary**

We used a new experimental technique to investigate the result of a meteoroid impact into an evaporitic deposit on Mars. Although two-stage light gas guns are ideal projectile launchers, the dirty gas from the gun has been a long-standing limitation of this technique that so far greatly complicated analysis of the vapors that are generated due to such impacts. Our new method overcomes this limitation and allows us to measure impact-generated vapor from evaporitic minerals. We detected NaCl vapor from halite and water vapor from gypsum at velocities lower than the typical impact velocities onto Mars. This suggests that volatile release from ancient dry lakebeds has periodically occurred throughout Martian history, due to stochastic meteoroid impacts. The nature of perchlorates and Ca-sulfates found on Mars can be interpreted as the result of hypervelocity impacts onto dry lakebeds rich in evaporitic minerals. **(141 words)**

## 1. Introduction

The presence of evaporitic minerals, such as halite (NaCl) and gypsum ($CaSO_4 \cdot 2H_2O$), on a planet can be interpreted as evidence that wet conditions existed in the planet's geological history. For example, sedimentation of evaporites is an inevitable outcome when a lake evaporates under playa-like conditions (e.g., Komatsu et al., 2007). Minerals interpreted to be of evaporitic origin (i.e., sulfates and chlorides) have been found on Mars (e.g., McLennan et al., 2005; Osterloo et al., 2008; Davila et al., 2011). Here, we consider the situation when hypervelocity impacts occur into ancient dry lakebeds rich in evaporitic minerals. Indeed, there appear to be impact craters in chloride-bearing deposits on Mars (Osterloo et al., 2010). The results of an impact into an evaporitic deposit have a number of interesting and diverse implications, including the promotion of perchlorate ($ClO_4^-$) formation and transformation of gypsum into anhydrite.

Recently, enrichment of chlorine (Cl) in the Martian surface with respect to bulk Mars has been detected, based on both the composition of Martian meteorites and remote sensing observations by a gamma ray spectrometer (GRS) onboard the Mars Odyssey (e.g., Lodders & Fegley, 1997; Rao et al., 2002; Keller et al., 2006). The Phoenix lander and Curiosity rover have detected perchlorate ($ClO_4^-$) in Martian soil at high and low latitude, respectively, which is one of the expressions of this excess Cl (e.g., Hecht et al., 2009; Kounaves et al., 2014). This implies that perchlorates are widespread over the entire surface of Mars (Clark & Kounaves, 2016). Given that perchlorate is a powerful oxidant in the surface environment, the distribution and behavior of Cl are important in understanding the survivability of organics on Mars. If chlorine-bearing minerals coexist with silicate minerals, perchlorates can be efficiently produced via oxidative photochemistry by ultraviolet radiation (<280 nm; UV-C; Carrier and Kounaves, 2015) and galactic cosmic rays (GCR; Wilson et al., 2016), and via high-temperature plasma chemistry by electrical discharges in dust storms (Wu et al., 2018). Recently, a model combining these two potential scenarios has also been proposed (Martínez-Pabello et al., 2019). The remaining question is how a sufficient amount of chlorine-bearing minerals was supplied to Martian soil, which initially contained a similar amount of Cl as bulk Mars. If shock-induced vaporization of halite commonly occurs during impact events, then hypervelocity impacts might have promoted extensive lateral transport of Cl-bearing minerals from halite deposits to surrounding Martian soils.

In this study, we also focus on shock-induced devolatilization of gypsum. Although known large-scale gypsum deposits on Mars are limited to the gypsum dunes of Olympia Undae near the North Pole region (Fishbaugh et al., 2007; Ehlmann and Edwards, 2014), the CheMin X-ray diffraction (XRD) instrument onboard Curiosity found the dehydrated forms of Ca-sulfates, basanite ($CaSO_4 \cdot 0.5H_2O$) and anhydrite ($CaSO_4$), as well as gypsum, in Gale Crater (e.g., Brake et al., 2013; Vaniman et al., 2018). The limited global-scale gypsum deposits and non-equilibrium nature of the sulfates in Gale Crater suggest that shock-induced water loss from gypsum has been an important geological process on Mars. In addition, shock-induced water loss from gypsum has been proposed as a new shock indicator (e.g., Bell and Zolensky, 2011).

To obtain insights into such processes, impact experiments on evaporite targets are desirable. For instance, the vaporization/devolatilization thresholds of evaporites are important to understand the thermodynamic response of evaporitic minerals to impacts. Two-stage light gas guns have been used to investigate impact-driven processes, because they can accelerate a macroscopic projectile (>0.1 mm in diameter) at room temperature up to velocities of several km $s^{-1}$ (e.g., Kurosawa et al., 2012). To avoid chemical contamination from the gases used for the operation of these guns, most previous studies have been conducted in closed systems (e.g., Boslough et al., 1982; Furukawa et al., 2008; Martins et al., 2013), in which samples are completely covered by containers. In contrast, all natural impacts occur in open systems. The thermodynamic path during decompression from a shocked to reference state in a closed system is significantly different from that in an open system because: (1) fast cooling of the shocked materials due to adiabatic expansion is prevented due to the limited volume of the container (e.g., Ivanov & Deutsch, 2002; Kurosawa et al., 2012); and (2) the total pressure in the closed system is supported by the mechanical strength of the container. Given that the mechanical strength of a container made of stainless steel reaches ~1 GPa (e.g., Nakazawa et al., 2005), the phase changes and chemical reactions proceed under high-pressure and low-temperature conditions (i.e., a low-entropy state) in a closed system. This situation is markedly different from natural impacts. Thus, impact experiments in an open system are essential to accurately understand impact-driven phenomena.

In a previous study, we developed an experimental technique using two-stage light gas guns (Kurosawa et al., 2012). This method allows us to conduct mass spectrometric analysis of gases generated in an open system using light gas guns. The method can prevent chemical contamination from the gun (Kurosawa et al., 2012; Ishibashi et al., 2017). However, the procedure requires use of a plastic diaphragm, which leads to destruction of weak projectiles, such as fused quartz (Kurosawa et al., 2012), and chemical contamination by trace amounts of carbon (ppm levels) due to carbon-bearing vapor produced by the penetration of the plastic diaphragm. This limitation prevents further applications of this method, such as accurate detection of incipient vaporization/devolatilization and a detailed investigation of complex organic species in impact-generated vapor. Thus, the objective of this study was to develop a system to investigate shock vaporization/devolatilization in an open system using a two-stage light gas gun without a diaphragm.

## 2. Experiments

We describe the detail of the experimental system developed in this study and experimental condition in Section 2.1. The experimental procedure are described in Section 2.2.

### 2.1. Experimental system and condition

We used the gun system developed at the Planetary Exploration Research Center of Chiba Institute of Technology (PERC/Chitech), Japan (Kurosawa et al., 2015), which has two new automatic gate valves installed in the gun system. Figure 1a shows a schematic diagram of the system.

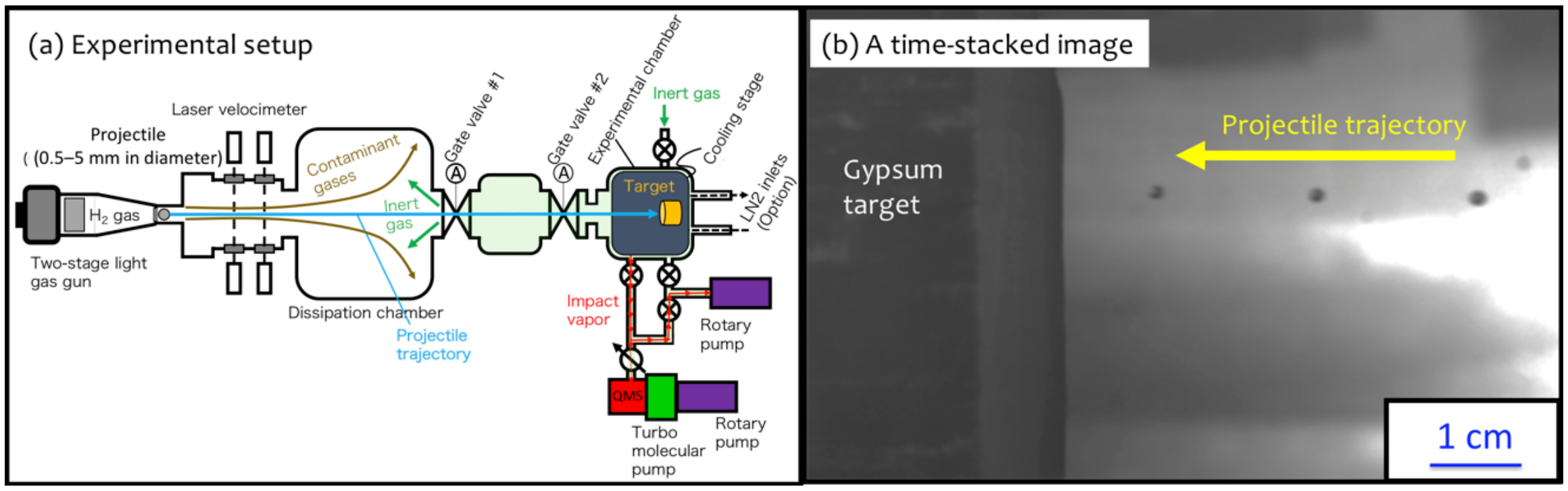

**Figure 1.** (a) Schematic diagram of the experimental system. (b) A time-stacked image of a flying projectile prior to an impact. The time interval between images is 4 μs.

Natural samples of halite and gypsum were used as targets, which were shaped as blocks. The masses of the halite and gypsum blocks were ~1 and ~0.6 kg, respectively. The halite targets are polycrystalline aggregates of randomly oriented crystals. We used a polycrystalline form of gypsum that is referred to as satin spar gypsum. The gypsum was cut into several blocks and all impacts were performed parallel to the *c*-axis. Given the axial compressibility of gypsum is almost isotropic up to 4 GPa (Comodi et al., 2008), the effects of crystal orientation on shock-induced devolatilization are likely to be insignificant. The chemical and mineralogical compositions of the samples are summarized in Supplementary Information Text S1. We also used a natural basalt block as a target in a control experiment. Spheres made of oxides (i.e., fused quartz and $Al_2O_3$) were used as projectiles. The projectile diameters were set to 2.0 and 1.5 mm for the halite and gypsum, respectively. We used the smaller projectile in the experiments using gypsum targets because the sizes of the gypsum blocks were smaller than those of the halite blocks. A nylon-slit sabot (Kawai et al., 2010) was used to accelerate the projectiles. The impact velocity was varied from 1.9 to 7.2 km/s. The peak pressures at the impact point were estimated to be 10–110 GPa by the one-dimensional impedance method (e.g., Melosh, 1989). Hereafter, the peak pressure is referred to as the 1-D pressure $P_{1D}$. The shock Hugoniot parameters were taken from previous studies (Marsh, 1980; Melosh, 1989; Ahrens & Johnson, 1995; Trunin et al., 2001) and the values actually used are summarized in Supplementary Information Text S2. The experimental chamber was pressurized to an equilibrium pressure determined by the balance between He gas injection and the evacuation by the pump. This prevented the intrusion of the contaminant gases from the gun as described in Section 2.2. The equilibrium pressure was set to 500 Pa. The target blocks were set on the target stage with a small tilt to minimize the damage to valve #2 by fast ejecta moving back in the direction of the projectile trajectory, which is frequently observed in cratering processes under the strength-dominated cratering regime (e.g., Hoerth et al., 2013). The impact angle was 80° from the target surface. Thus, the calculated $P_{1D}$ values are slightly overestimated in this study.

### 2.2. Experimental procedure

Prior to a shot, valves #1 and #2 are closed and opened, respectively. The gun side uprange from valve #1 is evacuated and the green shaded region is filled with a chemically inert gas, such as helium (He). A gas flow is induced using a rotary pump. To maintain a constant ambient pressure in the experimental chamber, the inert gas is continuously introduced into the experimental chamber with the same gas flux as the evacuation. Using a pulse generator, signals with different time delays are inputted to valves #1 and #2 and the two-stage light gas gun. At the start, valve #1 is opened, leading to the expansion of inert gas into the dissipation chamber due to the strong pressure gradient. The gun is then operated and a projectile is launched towards the target placed in the experimental chamber. The contaminant gas following the projectile is not able to intrude into the experimental chamber due to the outflow. In contrast, the solid projectile is able to penetrate into the outflow and impact the target. Since the time required for projectile acceleration is much shorter than the duration of the outflow, the impact occurs under a pressure that is ≥90% of the initial ambient pressure (see Supplementary Information Text S3.1). Immediately after the impact, valve #2 is closed to keep the impact-generated vapor in the experimental chamber. Subsequently, the gas generated by the impact is introduced into a quadrupole mass spectrometer (QMS, Pfeiffer vacuum, Prisma plus QMG220) by the inert gas flow produced by the rotary pump. This experimental method is termed the “two-valve method”. Figure 1b shows a time-stacked image of a fused quartz projectile immediately before the impact, clearly showing that the projectile was completely intact before the impact. Thus, the two-valve method allows us to use any combination of projectile and target.

To detect water vapor from gypsum, we used a cold trap placed at the bottom of the target stage to reduce the background water vapor in the chamber. We used liquid nitrogen ($N_2$) as a refrigerant. The temperature of the gypsum target prior to a shot was monitored with a thermocouple. The typical target temperature prior to a shot was ~250 K, which is intermediate between the average surface temperature (~210 K) and maximum diurnal temperature near the equator (~290 K) on Mars.

## 3. Results

Experimental results on halite and gypsum are presented in Sections 3.1 and 3.2, respectively, along with the mass spectrometry results on the shock-generated gases. The ion current for the mass number $M/Z = i$ is denoted as $I_i$. We also conducted another series of impact experiments using natural calcite samples to characterize the experimental system (Supplementary Information Text S3). We confirmed that the intrusion of contaminant gas from the gun was reduced to ~1 μg, which corresponds to 0.01–0.1 wt.% of the projectile mass.

### 3.1. Shock vaporization of halite

Figure 2a shows a typical example of the temporal variations of the ion currents at selected mass numbers. We detected a sudden rise in $I_{58}$ ($Na^{35}Cl^+$) and $I_{70}$ ($^{35}Cl_2^+$) after the impact, although the peak current ratios $I_{58}/I_4$ and $I_{70}/I_4$ were only 0.1–1.0 ppm. Although we only show the ion currents of $^{35}Cl$-bearing species, we observed similar trends for species containing $^{37}Cl$, such as $I_{60}$ ($Na^{37}Cl^+$), $I_{71}$ ($^{36}Cl^{37}Cl^+$), and $I_{72}$ ($^{37}Cl_2^+$). The $I_i$ to $I_4$ ($He^+$) ratio approximates the partial pressure of species $i$ in the experimental chamber, because the total pressure in the chamber was mostly supported by He (Supplementary Information Text S3.3). We also detected a rise in $I_{18}$ ($H_2O^+$), possibly from adsorbed water due to the highly hygroscopic nature of halite. The rise in $I_2$ ($H_2^+$) is likely to be caused by the cracking of $H_2O$ by electron impacts in the QMS. This rise was not due to the intrusion of contaminant gas from the gun, because the rise in $I_2$ in the experiments on natural calcite targets was much smaller than those in the halite experiments (see Supplementary Information Text S3.2).

Figure 3a shows the temporal variations in $I_{58}/I_4$ at different shock pressures. We found that halite initiates vaporization between 18 and 31 GPa.

### 3.2. Shock-induced water release from gypsum

Figure 2b is the same as Fig. 2a except that gypsum was the target. The cold trap allowed us to reduce the background level of $I_{18}$ down to ~1/30, as compared with the case of the halite experiments (Fig. 2a). Shock-generated water vapor was clearly detected. In contrast, sulfur-bearing gases, such as SO, $SO_2$, and $SO_3$, did not clearly increase after the impact even at 7.2 km/s. The estimated 1-D pressure of this shot reaches $P_{1D}$ = 114 GPa. Note that the time when $I_{18}$ becomes the maximum was delayed up to ~50 s after the time at the impact.

Figure 3b shows the same as Fig. 3a except that the target was gypsum and $I_{18}/I_4$ is shown. We observed increases in $I_{18}/I_4$ even at 2.0 km/s. In contrast, $I_{18}/I_4$ remained at the background level in the blank experiment, which used a basalt block at the same impact velocity of 2.0 km/s, suggesting that gypsum experienced water loss at $P_{1D}$ <11 GPa. In the control experiment, a fused quartz projectile was accelerated to 2.0 km/s. Thus, we can rule out the possibility that the detected water vapor in the gypsum experiment at the lowest impact velocity (2.0 km/s) comes from the desorption of water vapor from the chamber wall/floor or the projectile.

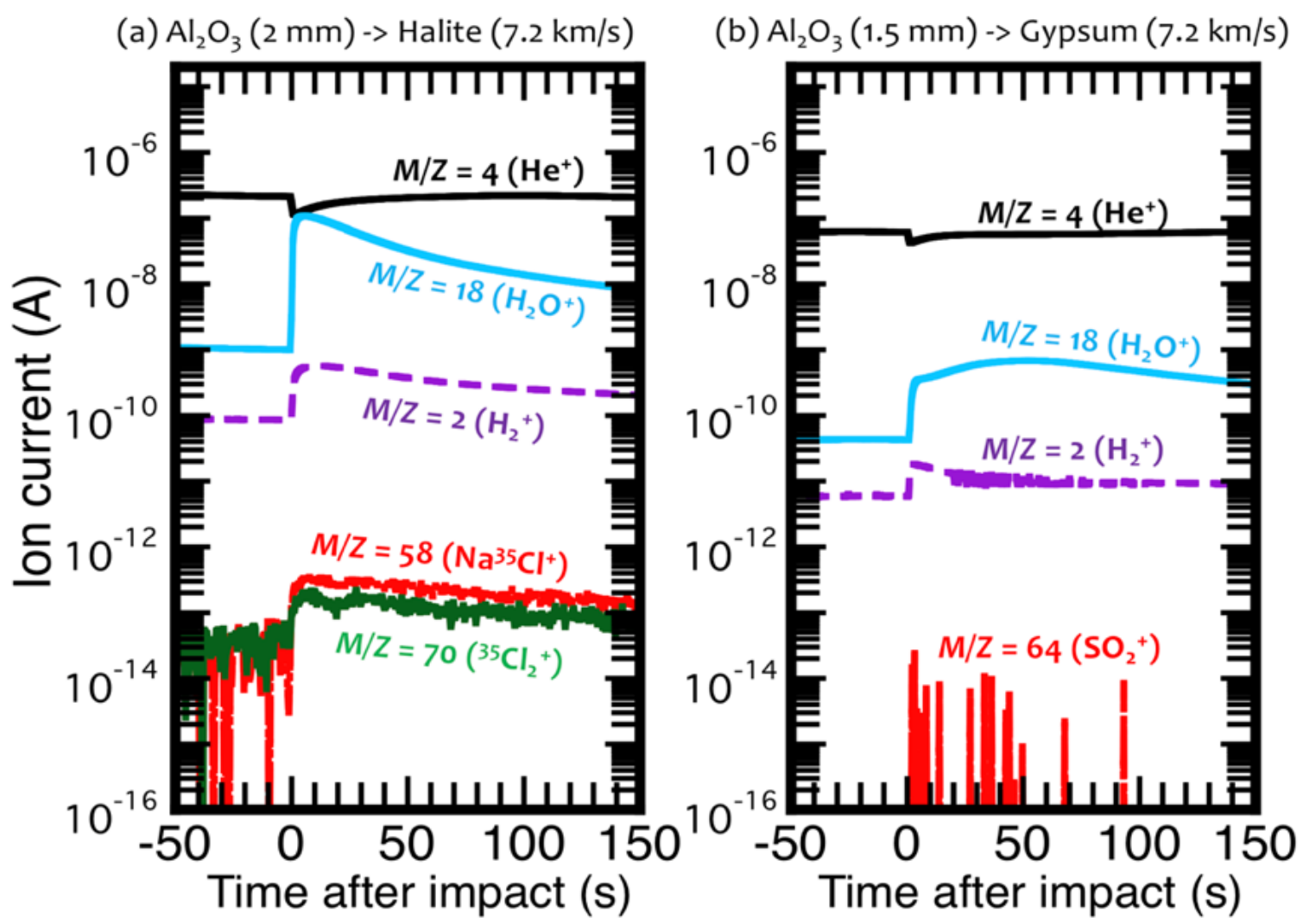

**Figure 2.** Time variations of the ion currents for the selected species for (a) halite and (b) gypsum. The projectile diameters and impact velocities of the shots are indicated on top of the figure.

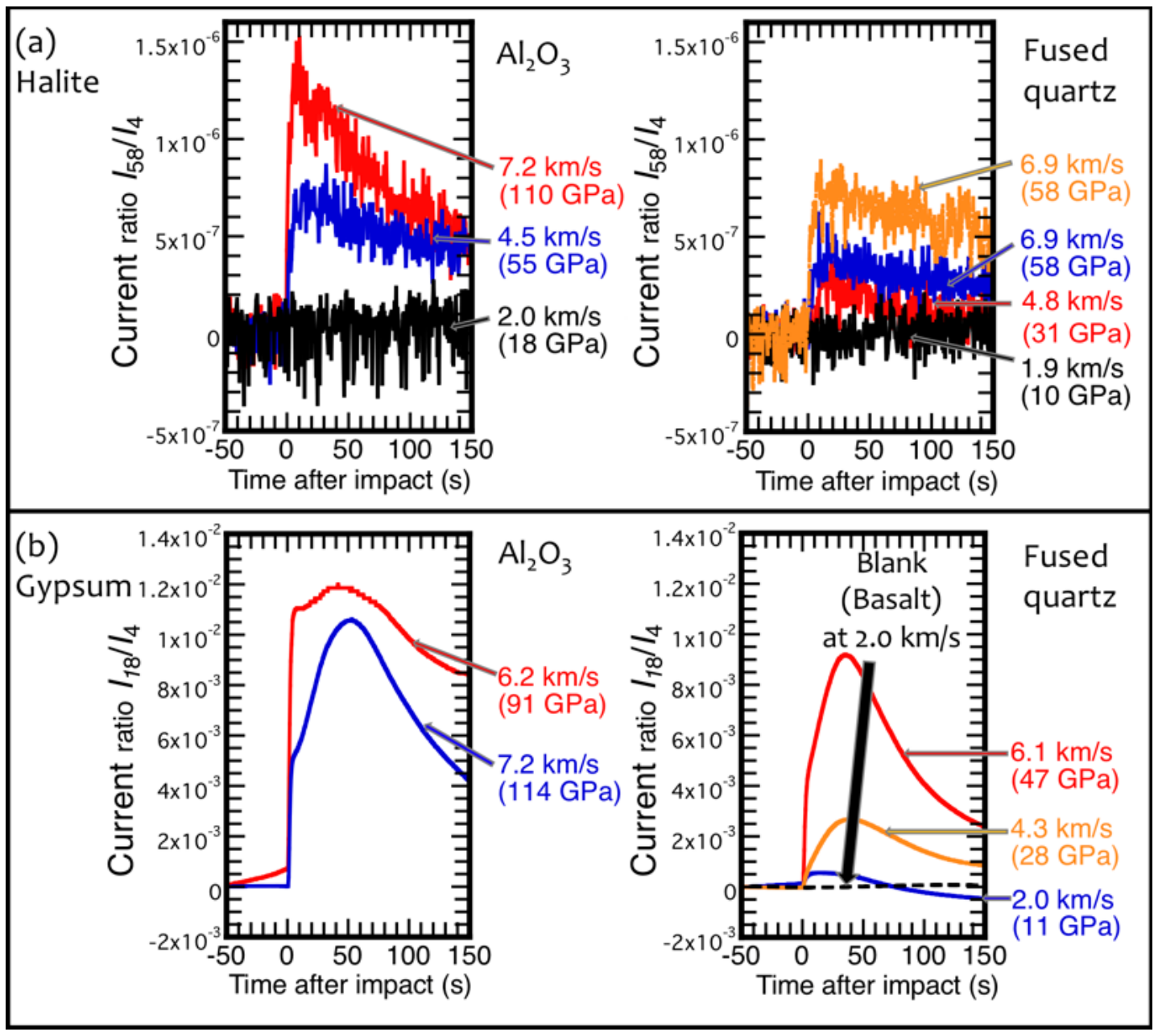

**Figure 3.** Time variations of the current ratios $I_{58}/I_4$ ($Na^{35}Cl^+/He^+$, a) and $I_{18}/I_4$ ($H_2O^+/He^+$ b). (a) The results for the (a) halite and (b) gypsum targets are shown. The results for $Al_2O_3$ and quartz projectiles are displayed at the left and right panels, respectively. The rises in $I_{58}/I_4$ and $I_{18}/I_4$ corresponds to the generation of NaCl and water vapor, respectively.

## 4. Discussion

In our experimental system, accurate quantitative measurements of the gases produced are not available due to gas escape from the chamber through valve #2 prior to its closure (see Supplementary Information Text S3.5). As such, we mainly discuss the initiation of evaporite vaporization/devolatilization with increasing 1-D pressure in Section 4.1. The initiation can be determined by the presence or absence of sudden increases in the ion current ratios after the impacts. We then discuss the chemical composition of the vapor generated from gypsum in Section 4.2. The implications of our results are discussed in Section 4.3.

### 4.1. Vaporization/devolatilization threshold

We estimated the boiling point of pure NaCl to be ~1,200 K using a Gibbs energy minimization method (e.g., Gordon & McBride, 1994) at 500 Pa (see Supplementary Information Text S4). We used 1,200 K as an approximate temperature required for incipient vaporization of halite because we confirmed that the halite targets are mainly composed of NaCl (99.5 wt.%; Supplementary Information Text S1). According to a previous study (Ballirano & Melis, 2009), water release from gypsum in a vacuum occurs at ~370 K. If the post-shock temperatures $T_{post}$ after pressure release down to 500 Pa exceed the above temperatures (i.e., 1,200 K for halite and 370 K for gypsum), then vaporization/devolatilization should occur (e.g., Ahrens & O'Keefe, 1972; Ivanov & Deutsch, 2002; Kurosawa et al., 2012). The ambient pressure in the experimental chamber before the closure of valve #2 decreases to ~350 Pa, although we assumed the reference pressure to be 500 Pa. If we use the minimum ambient pressure (350 Pa) as the reference pressure, the boiling point of NaCl becomes 1,150 K. This does not affect the conclusions described below. Figure 4 shows the post-shock temperature as a function of $P_{1D}$. We calculated $T_{post}$ in two steps as described as follows. First, we calculated the shock-induced entropy gain using the method proposed by Sugita et al. (2012). In this calculation, we used the shock Hugoniot parameters and Mie–Grüneisen EOS, and adopted the Dulong–Petit value for the constant isochoric specific heat. Second, we calculated the difference in temperature between the initial state and post-shock state by assuming isentropic release from the peak shock state (e.g., Senshu et al., 2002). We found that vaporization/devolatilization from the evaporitic minerals occurs at systematically lower $P_{1D}$ than expected from these simple thermodynamic estimates. This does not necessarily mean that the theoretical estimates are incorrect because of the reasons as follows. Although most previous experiments have reported the shock pressure required for incipient vaporization/devolatilization based on whether shock-generated gases were detected at a given shock pressure, our results suggest that local energy concentration due to the processes, such as jetting (e.g., Kieffer, 1977), shear banding (Kondo & Ahrens, 1983), and irreversible heating due to friction and plastic deformation (Kurosawa & Genda, 2018), largely contribute to the initiation of gas release with increasing $P_{1D}$. Consequently, experimental assessment of the shock pressure required for initiating gas release are complicated by disequilibrium energy partitioning into local regions and/or strength-induced irreversible heating.

We now consider the differences in spatial scale between experimental and natural impact events. The spatial scale of natural impacts is at least several orders of magnitude greater than that of laboratory-scale impacts, leading to a much lower cooling/decompression rate in natural impact events. Given that we can assume thermodynamic equilibrium and no heat production during pressure release (i.e., an adiabatic process), the decompression path on a pressure–temperature plane should be the same for natural and laboratory-scale impacts (e.g., Ahrens and O'Keefe, 1972), although the times required for complete decompression are quite different (e.g., Ishimaru et al., 2010). In this case, the simple thermodynamic estimates based on $T_{post}$ to determine the pressure required for incipient vaporization/devolatilization (i.e., water loss) can be applied to natural impacts.

The maximum degree of impact heating due to the local energy concentration caused by jetting and shearing would not be strongly dependent on the spatial scale, because it is determined by the local flow/stress field during an impact (e.g., Kieffer, 1977; Kondo and Ahrens, 1983). However, heating due to plastic deformation in pressure-strengthened rocks during pressure release (Kurosawa and Genda, 2018) is potentially a scale-dependent process, although this has not been investigated. Melosh and Ivanov (2018) noted that the yield stress of rocky materials also depends on the strain rate, although Kurosawa and Genda (2018) used a scale-independent strength model constructed from static failure data (e.g., Lundborg, 1968). A comparative study of laboratory/numerical impact experiments and natural samples that have experienced shock-induced metamorphism might provide a robust understanding of scaling effects on shock-induced vaporization/devolatilization. Such an investigation is the beyond the scope of the present study.

## 4.2. Stability of sulfur during impact shocks

Release of sulfur-bearing gases from the gypsum targets was not clearly detected even at $P_{1D}$ = 114 GPa (Fig. 2b). Nevertheless, the required shock pressure for incipient/complete sulfur release from anhydrite ($CaSO_4$) has been estimated to be 81±7 GPa/155±13 GPa (Yang et al., 1998). This large difference between gypsum and anhydrite suggests that the stability of sulfur during impact shocks depends on the initial form of the sulfur-bearing sedimentary rocks. As such, the extreme environmental perturbations caused by impact-driven sulfur release during the Cretaceous–Paleogene (K–Pg) impact event (e.g., Ohno et al., 2014) might not have occurred if the bedrock of the impact site was gypsum. Further impact experiments using gypsum and anhydrite, along with solid phase analyses of the solid residues, are necessary to confirm this possibility.

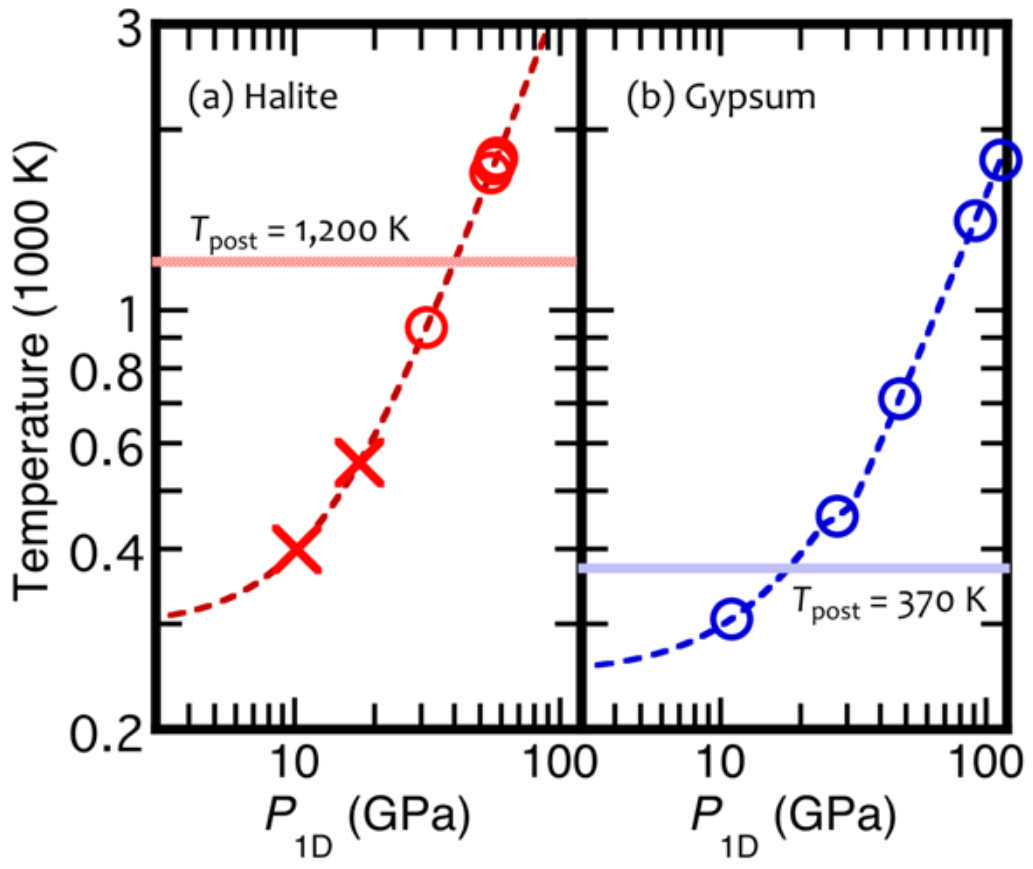

**Figure 4.** Post shock temperature for (a) halite and (b) gypsum as a function of the peak pressure at the impact point $P_{1D}$. The data points on the curves indicate $P_{1D}$ investigated in this study. The horizontal lines indicate the temperatures required for incipient vaporization and devolatilization (i.e., water loss) (Section 4.1). The circles indicate that the shock-generated gases were detected. The crosses represent that no gas release occurred.

4.3. Implications

Several mechanisms have been proposed that could lead to the lateral transport of Cl-bearing minerals (e.g., Carrier and Kounaves, 2015); i.e., transportation by impact ejecta, aqueous films, liquid runoff, or infiltration. We detected vaporization of halite at $P_{1D}$ = 18–31 GPa, although the peak $I_{58}/I_4$ value was only ~1 ppm. Given that these $P_{1D}$ values are easily achieved in natural impact events at ~4 km/s (Supplementary Information Text S2), impacts onto dry lakebeds at typical impact velocities on Mars would cause shock-induced vaporization of halite. Subsequent expansion of the vapor and condensation of fine NaCl particles would significantly enhance the lateral transport and reactivity of Cl on Mars. Thus, shock-induced vaporization might have been responsible for the production of perchlorates over the Martian surface in the past 3 Gyr, since the hypothesized time of formation of dry lakebeds linked to fluvial activity (e.g., Ramirez and Craddock, 2018). Given the stochastic nature of natural impacts, the impact-driven transport of Cl from halite lakebeds potentially explains the spatial heterogeneity of Cl on the Martian surface observed by GRS.

The mass of water vapor $M_{H2O}$ produced from gypsum can be roughly estimated using the method described in Supplementary Information Text S3.4. $M_{H2O}$ is at least 0.1–1 times the projectile mass at a $P_{1D}$ of 10–110 GPa, even after taking into account the effect of gas escape from the experimental chamber (see Supplementary Information Text S3.5). A large volume of anhydrite equivalent to the volume of an impactor might be produced after an impact. Consequently, impact-driven water loss from gypsum could explain the coexistence of gypsum and anhydrite at the Gale crater (e.g., Vaniman et al., 2018), which might have been caused by the events that formed the impact craters found superimposed onto the Gale crater (e.g., Le Deit et al., 2014). Shocked gypsum ejecta, i.e., anhydrite, might have been deposited at/onto the Curiosity sampling sites. In addition, our experiments support the hypothesis that gypsum can serve as a shock indicator, given its high sensitivity at relatively low $P_{1D}$ (~10 GPa) (e.g., Bell & Zolensky, 2011)

## 5. Conclusions

We developed a new experimental method to investigate shock-induced vaporization/devolatilization in an open system. This method allows us to use a two-stage light gas gun to investigate the nature of shock-driven phase changes and chemical reactions. Vaporization/devolatilization of evaporitic minerals (halite and gypsum) that are found on Mars were investigated with this method. The shock pressures required for the incipient vaporization of halite and devolatilization of gypsum were estimated to be 18–31 and <11 GPa, respectively. These values are systematically lower than theoretical thermodynamic estimates, suggesting that the initiation of shock-induced gas release are determined by local energy concentrations caused by jetting and shear banding, and/or strength-induced irreversible heating. Hypervelocity impacts might have promoted the formation of perchlorates in Martian soil throughout Martian history and be responsible for the coexistence of different forms of Ca-sulfates in the same impact crater.

## Acknowledgments, Samples, and Data

This work was supported by ISAS/JAXA as a collaborative program with its Hypervelocity Impact Facility. We thank Sunao Hasegawa for assistance with the development of our experimental system. We thank Takashi Mikouchi for providing the natural gypsum sample. We appreciate helpful comments by Kojiro Suzuki and Kazuhisa Fujita on the design of the two-valve method. We also thank Reika Yokochi for helpful suggestions about the reduction of background levels of water vapor in our system. We appreciate useful discussions at a workshop on planetary impacts held in 2018 at Kobe University, Japan. The chemical compositions of the rock samples were analyzed by Satoshi Takenouchi at NIMS. We also thank Mattihias Ebert and Christopher Hamann for their constructive reviews that helped greatly improve the manuscript, and Andrew J. Dombard for handling the manuscript as an Editor. KK is supported by JSPS KAKENHI Grant Numbers 23840057, 25871212, 17H01176, 17H02990, 17H01175, 17K18812, 18HH04464, and 19H00726 and by the Astrobiology Center of the National Institute of Natural Sciences, NINS (AB261014 and AB281026). Supporting information can be found in the online version of the manuscript. The data supporting the figures are available from the Academic Repository of Chiba Institute of Technology (http://id.nii.ac.jp/1196/00000229/).

**Supporting Information for**

# Shock vaporization/devolatilization of evaporitic minerals, halite and gypsum, in an open system investigated by a two-stage light gas gun

Kosuke Kurosawa[1]*, Ryota Moriwaki[1], Goro Komatsu[1,2], Takaya Okamoto[3], Hiroshi Sakuma[4], Hikaru Yabuta[5] and Takafumi Matsui[1]

[1] Planetary Exploration Research Center, Chiba Institute of Technology, 2-17-1, Tsudanuma, Narashinon, Chiba 275-0016, Japan.

[2] International Research School of Planetary Sciences, Università d'Annunzio, Viale Pindaro, 42, 65127 Pescara, Italy.

[3] Institute of Space and Astronautical Science, Japan Aerospace Exploration Agency, 3-1-1, Yoshinodai, Chuo-ku, Sagamihara, Kanagawa 252-5210, Japan.

[4] Research Center for Functional Materials, National Institute for Materials Science, 1-1, Namiki, Tsukuba, Ibaraki 305-0044, Japan.

[5] Department of Earth and Planetary Systems Science, Graduate School of Science, Hiroshima University, 1-3-1, Kagamiyama, Higashi–Hiroshima, Hiroshima 739-8526, Japan.

## Contents of this file

Text S1 to S4
Figures S1 to S11
Tables S1 to S4

## Introduction

This document includes sample descriptions (Text S1; Figs S1–S3; Tables S1–S3), the shock Hugoniot parameters used to calculate the shock pressure at the impact point in the main text (Text S2; Fig. S4; Table S4), the experimental system (Text S3.1–S3.6; Figs S5–S10), and the boiling temperature of NaCl under the experimental conditions of this study (Text S4; Fig. S11).

## Text S1. Chemical and mineralogical compositions of the samples

We confirmed that our natural samples are calcite, halite, and gypsum by X-ray diffraction (XRD; Ultima IV, Rigaku Corporation). The XRD measurements were performed with Cu Kα radiation of $\lambda$ = 0.154 nm at 40 kV, 30 mA, and a scan speed of 2°/min. The Miller indices of Bragg peaks were assigned on the basis of the American Mineralogist Crystal Structure Database (Downs & Hall-Wallace, 2003). However, some unknown peaks were also detected in the XRD measurements. Thus, we also determined the chemical compositions of the samples by inductively coupled plasma optical emission spectroscopy (ICP-OES; SPS3520UV-DD, Hitachi High-Tech Science Corporation) and a carbon–sulfur analyzer (Leco CS-444, LECO Corporation) at the National Institute for Materials Sciences (NIMS, Japan). We used two different pieces of each sample for the measurements. The chemical compositions of the samples are summarized in Tables S1 to S3.

*Calcite*

The XRD profile of calcite is shown in Fig. S1. Most Bragg reflections were assigned to calcite and single reflection of dolomite was also observed. The samples include a small amount of Mg (0.53 wt.%). Based on the atomic ratio of Ca to Mg, and the assumption that all the measured Ca and Mg are included as calcite ($CaCO_3$) and dolomite ($CaMg(CO_3)_2$), respectively, the fractions of $CaCO_3$ and $CaMg(CO_3)_2$ were estimated to be 96 and 4 wt.%, respectively. We also detected a small amount (0.2 wt.%) of sulfur in the calcite, suggesting that rare Ca and/or Mg sulfate minerals are present in the samples.

*Halite*

The XRD profile of salt is shown in Fig. S2. All detected Bragg reflections were assigned to halite. The samples are composed mainly of Na and Cl (99.5 wt.%). The impurities in the samples are K, Mg, and Ca.

*Gypsum*

The XRD profile of gypsum is shown in Fig. S3. Most Bragg reflections were assigned to gypsum. The Ca content was 23.3 wt.%. If it is assumed that all the measured Ca is included in the samples as gypsum ($CaSO_4•2H_2O$), then the mass fraction of $CaSO_4•2H_2O$ is estimated to be 100.1 wt.%, showing that the samples are pure gypsum.

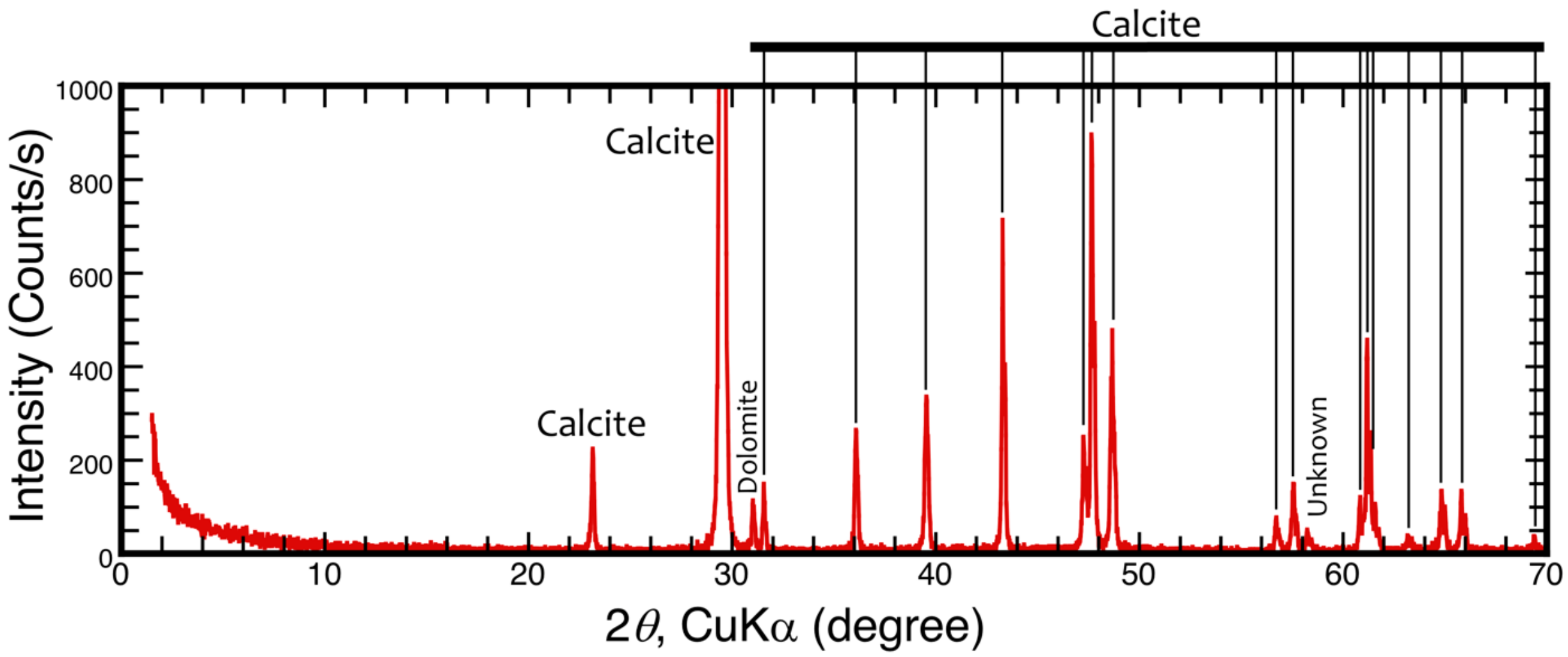

**Figure S1.** X-ray diffraction spectra of the natural calcite samples. The detected peaks are indicated in the figure. There are a few unknown peaks as indicated in the figure.

**Table S1.** Elemental composition of calcite.

| Sample No. | 1 | 2 | mean (wt.%) |
|---|---|---|---|
| Li | <0.001 | <0.001 | <0.001 |
| Na | <0.01 | <0.01 | <0.01 |
| Mg | 0.53 | 0.54 | 0.53 |
| Al | <0.01 | <0.01 | <0.01 |
| Si | <0.01 | <0.01 | <0.01 |
| K | <0.1 | <0.1 | <0.1 |
| Ca | 39.5 | 39.1 | 39.3 |
| Ti | <0.01 | <0.01 | <0.01 |
| Mn | <0.001 | <0.001 | <0.001 |
| Fe | <0.01 | <0.01 | <0.01 |
| Ni | <0.01 | <0.01 | <0.01 |
| S | 0.2 | 0.2 | 0.2 |
| C | 12.0 | 11.9 | 12.0 |
| | | | |
| Total | 52.2 | 51.7 | 52.0 |

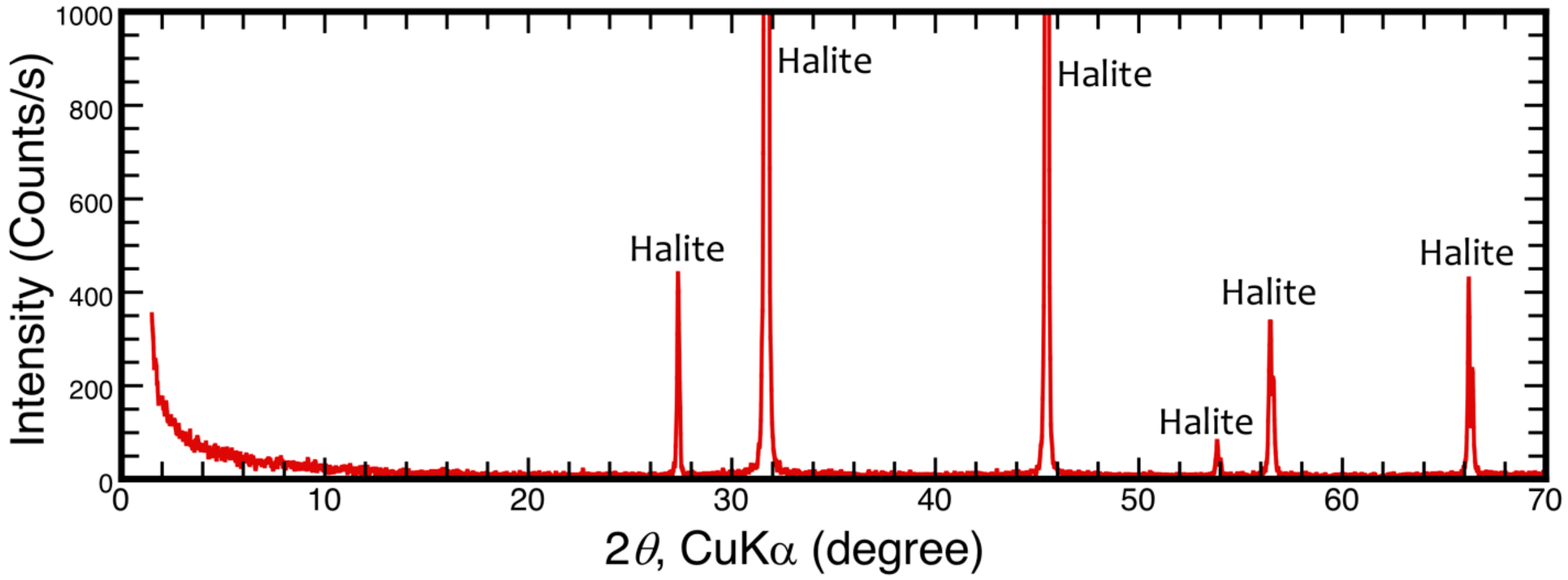

**Figure S2.** Same as Fig. S1 except that the XRD spectra of the natural halite samples are shown.

**Table S2**. Elemental composition of halite.

| Sample No. | 3 | 4 | mean (wt.%) |
|---|---|---|---|
| Li | <0.001 | <0.001 | <0.001 |
| Na | 39 | 39 | 39 |
| K | 0.1 | 0.1 | 0.1 |
| Mg | 0.01 | 0.02 | 0.02 |
| Ca | 0.04 | 0.06 | 0.05 |
| Rb | <0.01 | <0.01 | <0.01 |
| Fe | <0.01 | <0.01 | <0.01 |
| | | | |
| Cl | 60.3 | 60.7 | 60.5 |
| Total | 99.45 | 99.88 | 99.67 |

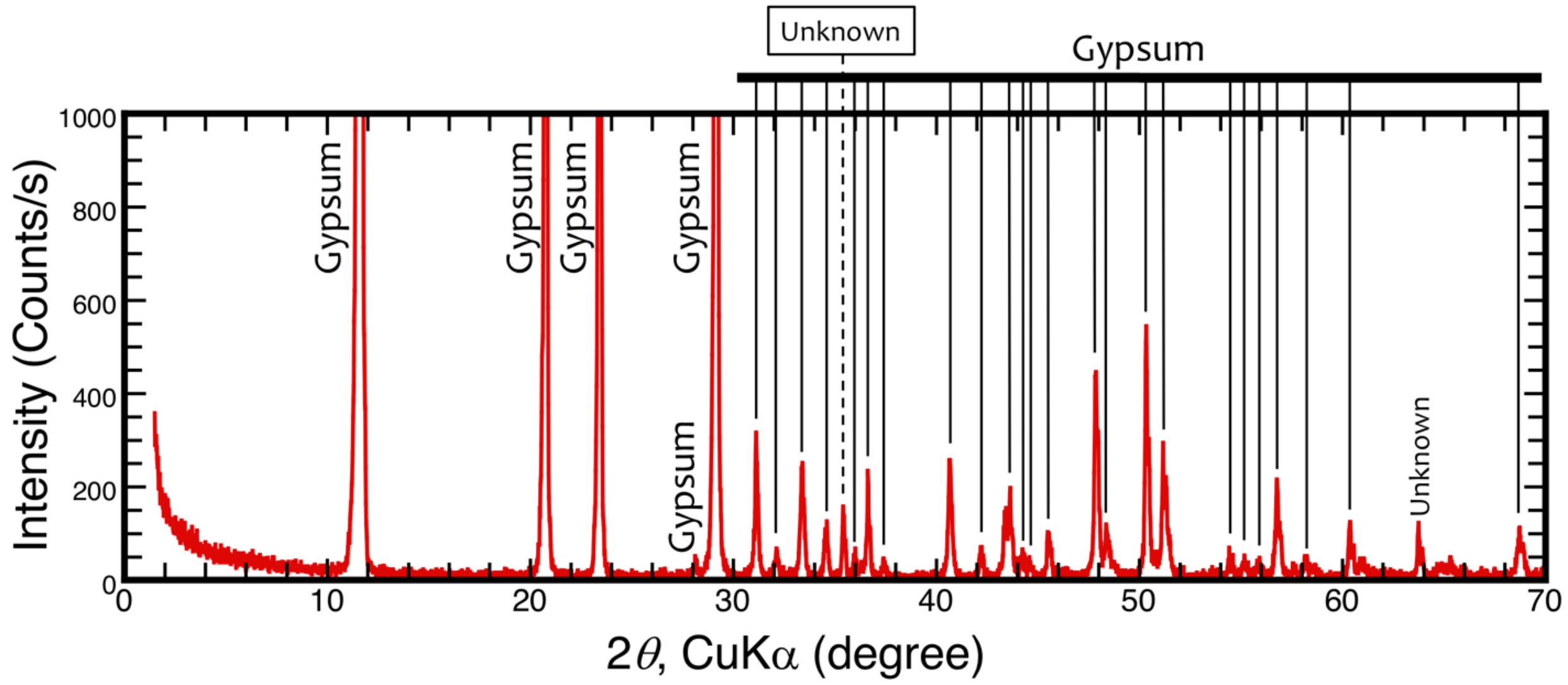

**Figure S3.** Same as Fig. S1 except that the XRD spectra of the natural gypsum samples are shown. There are a few unknown peaks as indicated in the figure.

**Table S3**. Elemental composition of gypsum.

| Sample No. | 1 | 2 | mean (wt.%) |
|---|---|---|---|
| Li | <0.001 | <0.001 | <0.001 |
| Na | <0.01 | <0.01 | <0.01 |
| Mg | <0.001 | <0.001 | <0.001 |
| Al | <0.01 | <0.01 | <0.01 |
| Si | <0.01 | <0.01 | <0.01 |
| K | <0.1 | <0.1 | <0.1 |
| Ca | 23.2 | 23.4 | 23.3 |
| Ti | <0.01 | <0.01 | <0.01 |
| Mn | <0.001 | <0.001 | <0.001 |
| Fe | <0.01 | <0.01 | <0.01 |
| Ni | <0.01 | <0.01 | <0.01 |
| S | 19.1 | 19.1 | 19.1 |
| Total | 42.3 | 42.5 | 42.4 |

**Text S2. Shock Hugoniot parameters and calculation of 1-D pressure**

We estimated the peak pressures at the impact point (i.e., the 1-D pressure) for each shot from the measured projectile speeds and one-dimensional impedance matching method (e.g., Melosh, 1989). The calculation employed the following quadratic equation: $V_s = C_0 + au_p + bu_p^2$, where $V_s$ is the shock velocity, $C_0$ is the bulk speed of sound, $a$ and $b$ are constants, and $u_p$ is the particle velocity. The $C_0$, $a$, and $b$ values are referred to as the "shock Hugoniot parameters" in this study. We list the parameters actually used in the calculation in Table S4.

Figure S4 shows the 1-D pressure $P_{1D}$ as a function of impact velocity. We used two different materials (i.e., fused quartz and $Al_2O_3$) as the projectile. The $Al_2O_3$ projectile yields a higher $P_{1D}$, because the shock impedance of $Al_2O_3$ is higher than that of fused quartz. Given that the impact velocity is known, we can estimate the 1-D pressure for each shot. The range of $P_{1D}$ values in this study is 10–110 GPa. For reference, we also calculated $P_{1D}$ for the impact of a basaltic projectile onto halite and gypsum targets (black dotted line in Fig. S4), in order to estimate $P_{1D}$ for natural impact events. We used the shock Hugoniot parameters for Kinosaki basalt taken from Sekine et al. (2008). Given that the shock impedance of corundum is much higher than that of Kinosaki basalt, the range of $P_{1D}$ values in this study is equivalent to velocities of up to ~9 km/s for natural impacts.

**Table S4**. Shock Hugoniot parameters used in this study.

| Material | Bulk density (g cm$^{-3}$) | Bulk speed of sound $C_0$ (km s$^{-1}$) | Constant $a$ | Constant $b$ (s km$^{-1}$) | Applicable range of $u_p$ (km s$^{-1}$) | Reference |
|---|---|---|---|---|---|---|
| Fused quartz[1] | 2.204 | 4.07 | 1.55 | 0.0 | 0.00–0.77 | Marsh (1980) |
| | | 5.07 | 0.00 | 0.0 | 0.77–2.25 | |
| | | 3.12 | 0.89 | 0.0 | 2.25–2.88 | |
| | | 0.76 | 1.71 | 0.0 | 2.88–3.19 | |
| Corundum[1] | 3.977 | 8.724 | 0.97535 | 0.0 | 0.56–5.38 | Marsh (1980) |
| Kinosaki basalt | 2.70 | 3.5 | 1.3 | 0.0 | 0.37–5.00 | Sekine et al. (2008) |
| Calcite[2] | 2.67 | 3.8 | 1.42 | 0.0 | | Melosh (1989) |
| Halite | 2.16 | 3.315 | 1.456 | –0.0219 | 0.59–11.05 | Trunin et al. (2001) |
| Gypsum | 2.28 | 2.80 | 1.95 | 0.0 | 0.85–1.72 | Ahrens & Johnson (1995) |
| | | 5.2 | 0.5 | 0.0 | 1.72–2.15 | |
| | | 2.49 | 1.79 | 0.0 | 2.15–4.06 | |

[1] We derived a stepwise linear relationship between shock speed and particle velocity from the data of Marsh (1980).

[2] We used a mean-fit over the full range of particle velocities, ignoring the solid–solid phase transition.

**Text S3. Characterization of the experimental system using calcite targets**

We conducted a series of impact experiments on natural calcite samples as mentioned in the main text. Calcite is the best material to explore the differences between the two-valve method developed in this study and previous experimental approaches (e.g., Kurosawa et al., 2012) because Kurosawa et al. (2012) used the same calcite samples as this study. In addition, there are numerous experimental data and theoretical estimates for the devolatilization of calcite (e.g., Boslough et al., 1982; Vizgirda and Ahrens, 1982; Lange and Ahrens, 1986; Martinez et al., 1995; Pierazzo et al., 1998; Ivanov and Deutsch, 2002; Gupta et al., 2002; Ohno et al., 2008; Kawaragi et al., 2009; Kurosawa et al., 2012; Bell, 2016). The two-valve method has some limitations (see Text S3.1 and S3.4) compared with the method proposed by Kurosawa et al. (2012). However, the two-valve method allows us to simulate natural impact phenomena almost perfectly, apart from with respect to scale, and quantitative measurement of the gas produced by the impacts is still challenging.

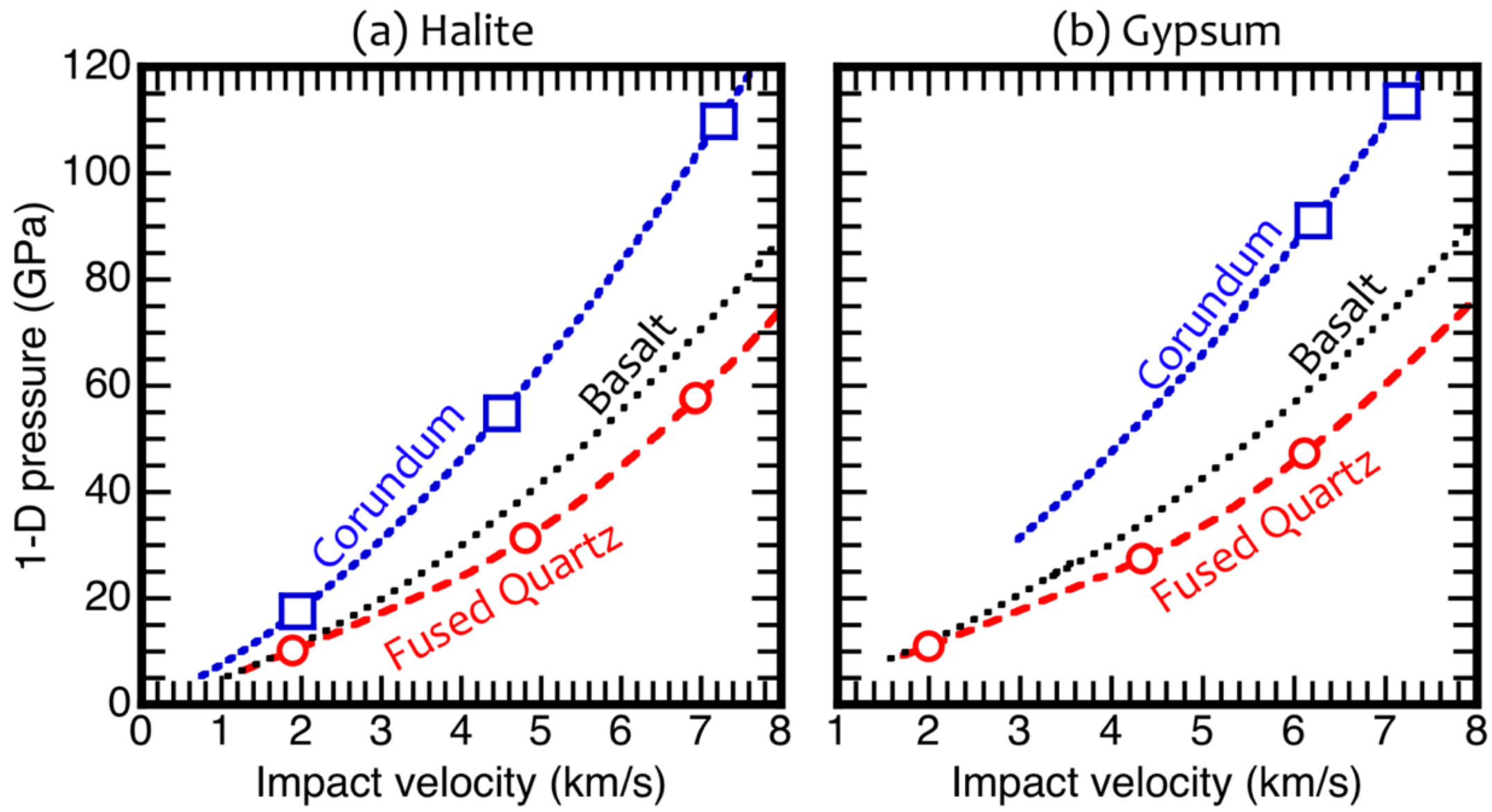

**Figure S4.** Calculated peak pressures at the impact point $P_{1D}$ at the measured impact velocities. The results for (a) halite and (b) gypsum targets are shown. We calculated $P_{1D}$ pertaining to two different projectiles, fused quartz (red open circles) and corundum (blue open squares). We also calculated $P_{1D}$ in the case of impact of a basaltic projectile onto halite and gypsum targets (black dotted line). The three lines on each figure show the results of the one-dimensional impedance match solutions. Given that the projectile speeds are measured, the $P_{1D}$ could be obtained by referencing to the lines.

*S3.1 Pressure variation during the experiments*

In the experiments with the two-valve method, the experimental chamber was not sealed by any valves, resulting in a fluctuation of the equilibrium total pressure in the chamber (see Fig. 2 in the main text). We examined the fluctuation in the ambient pressure using a vacuum gauge. Figure S5 shows the temporal change in the total pressure of the experimental chamber in the case of a He pressure of 500 Pa. Although the total pressure decreased gradually with time after valve #1 was opened down to ~350 Pa, the total pressure at the impact was still 90% of the initial value. The typical fluctuation in the pressure of the system was <30%.

*S3.2 Temporal variation of the ion currents*

Figure S6 shows a typical example of gas measurements in the calcite experiments. The impact velocity was 7.2 km/s, which is the highest in this series of experiments. Figure S6 is the same as Figs 2 and 3 in the main text, except that the selected mass numbers were optimized for calcite. The ion current for the mass number $M/Z = i$ is denoted as $I_i$. We clearly observed a sudden increase in $I_{44}$ ($CO_2^+$), showing that the shock-generated $CO_2$ from the calcite was detected by the two-valve method. The overall trend in the temporal variation of the ion currents is similar to that obtained by Kurosawa et al. (2012). Since the calcite samples did not contain water, the change in $I_{18}$ ($H_2O^+$) is much smaller than for the halite and gypsum targets (Fig. 2 in the main text). This implies that the observed water vapor in the halite experiments comes from the adsorbed water on the surface of the halite due to its highly hygroscopic nature.

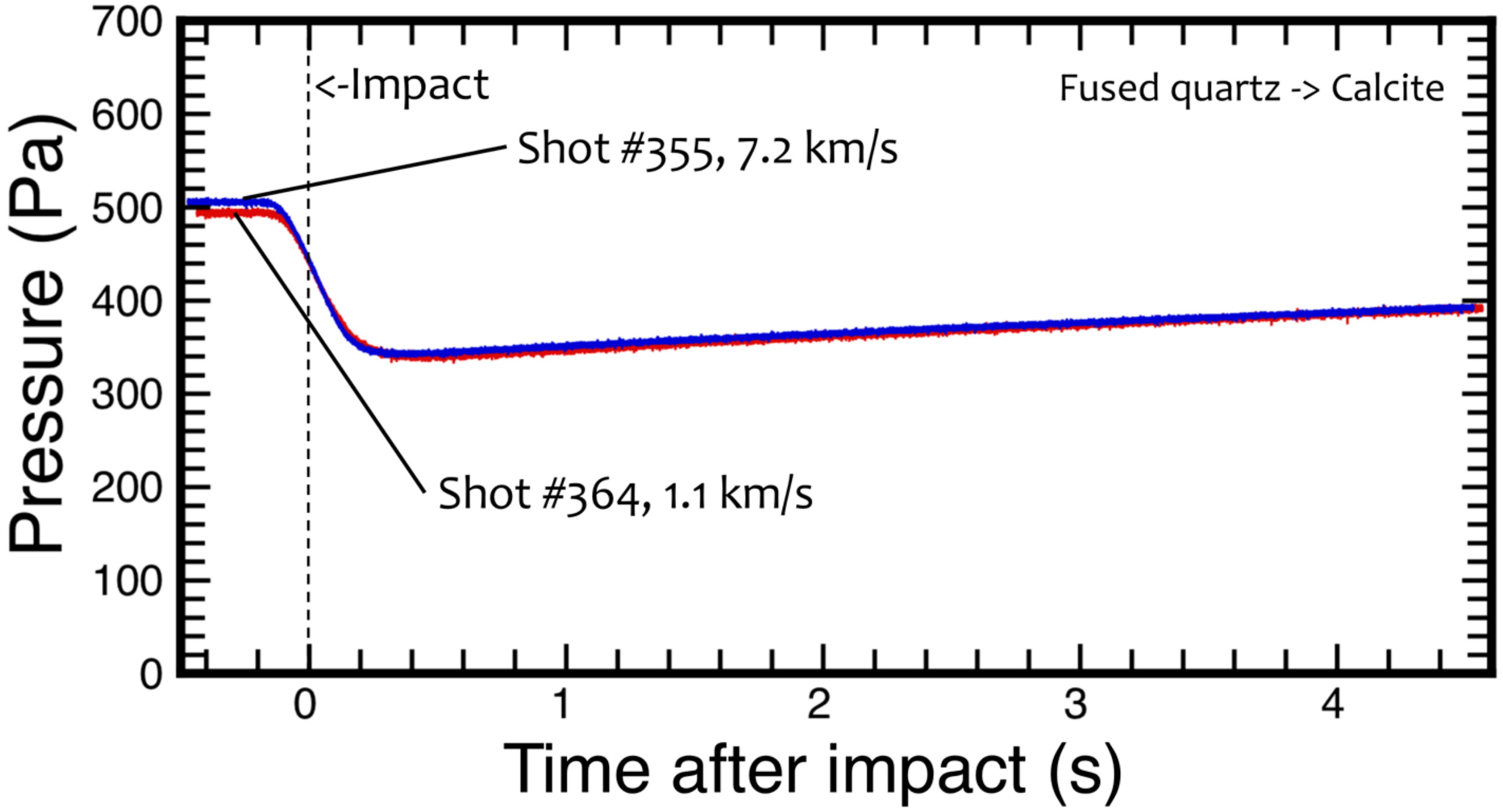

**Figure S5.** Time variation of the total pressure in the experimental chamber. The impact occurred at ~400 ms after the valve open. The results from different two shots at the highest and lowest projectile speeds performed in the calcite experiments. The total pressure at the impact was still ~90% of the initial total pressure.

*S3.3 Sensitivity of the quadrupole mass spectrometer*

We conducted a calibration experiment using standard $H_2$ and $CO_2$ gases to investigate the sensitivity of the quadrupole mass spectrometer (QMS) used in this study. We produced gas mixtures with different molar fractions of $H_2$ or $CO_2$ by dilution with He gas. The molar fractions of $H_2$ and $CO_2$ were controlled by changing the ratio of the partial pressure of $H_2$ or $CO_2$ to He. The gas mixtures were continuously introduced into the QMS. Figure S7 shows the pressure ratio plotted versus the measured ion current ratio. Our QMS system exhibits good linearity. Although the ion current ratio does not perfectly coincide with the pressure ratio, the difference is within a factor of 5 for both $H_2$ (a) and $CO_2$ (b). Thus, the ion current ratio can be approximated as the pressure ratio as noted in Section 3.1 of the main text. However, the ion current ratio to $I_4$ inevitably includes a ~30% uncertainty, due to the pressure fluctuation in the partial pressure of He in the experimental chamber as described in Text S3.1.

*S3.4 Hydrogen intrusion*

The most important limitation of conventional open-system experiments is the intrusion of contaminant gases from the gun into the experimental chamber as noted in Section 1 of the main text. We assessed the degree of contamination in the two-valve method using the increase in $I_2$ ($H_2^+$), because $H_2$ is the major contaminant gas from our gun used to accelerate the projectiles. Figure S8 shows $I_2/I_4$ as a function of time after the impact at different impact velocities. The peak values of $I_2/I_4$ range from 50 to 130 ppm at impact velocities of 3.3 to 7.2 km/s. No clear correlation between the peak values and impact velocities was observed. Since the QMS sensitivity for $H_2$ is known from Fig. S7a, we estimated the total mass of the intruded hydrogen $M_{H2}$ from $I_2/I_4$ (Fig. S8), the chamber volume ($V_{ch} \sim 40 \times 10^{-3}$ m$^3$), gas temperature ($T_{gas} \sim 300$ K), and the ideal gas EOS. We assumed that the intruded $H_2$ uniformly mixed with the He gas in the chamber. The intruded mass $M_{H2}$ is given by:

$$M_{H2} = \frac{\mu_{H2} P_{H2} V_{ch}}{R T_{gas}} \quad (S1)$$

where $R = 8.314$ J K$^{-1}$ mol$^{-1}$, $\mu_{H2} = 2$ g mol$^{-1}$, and $P_{H2}$ are the gas constant, molecular weight of $H_2$, and partial pressure of $H_2$, respectively. We can estimate $P_{H2}$ from the peak value of the ion current ratio and QMS sensitivity. The estimated $M_{H2}$ is ~1 $\mu$g, which corresponds to 0.01–0.1 wt.% of the projectile mass (4–17 mg, depending on the density and diameter) and 5 ppm of the mass of $H_2$ gas used to operate the gun (200 mg). We used $I_2/I_4 = 10^{-4}$ as the peak value in the above calculation.

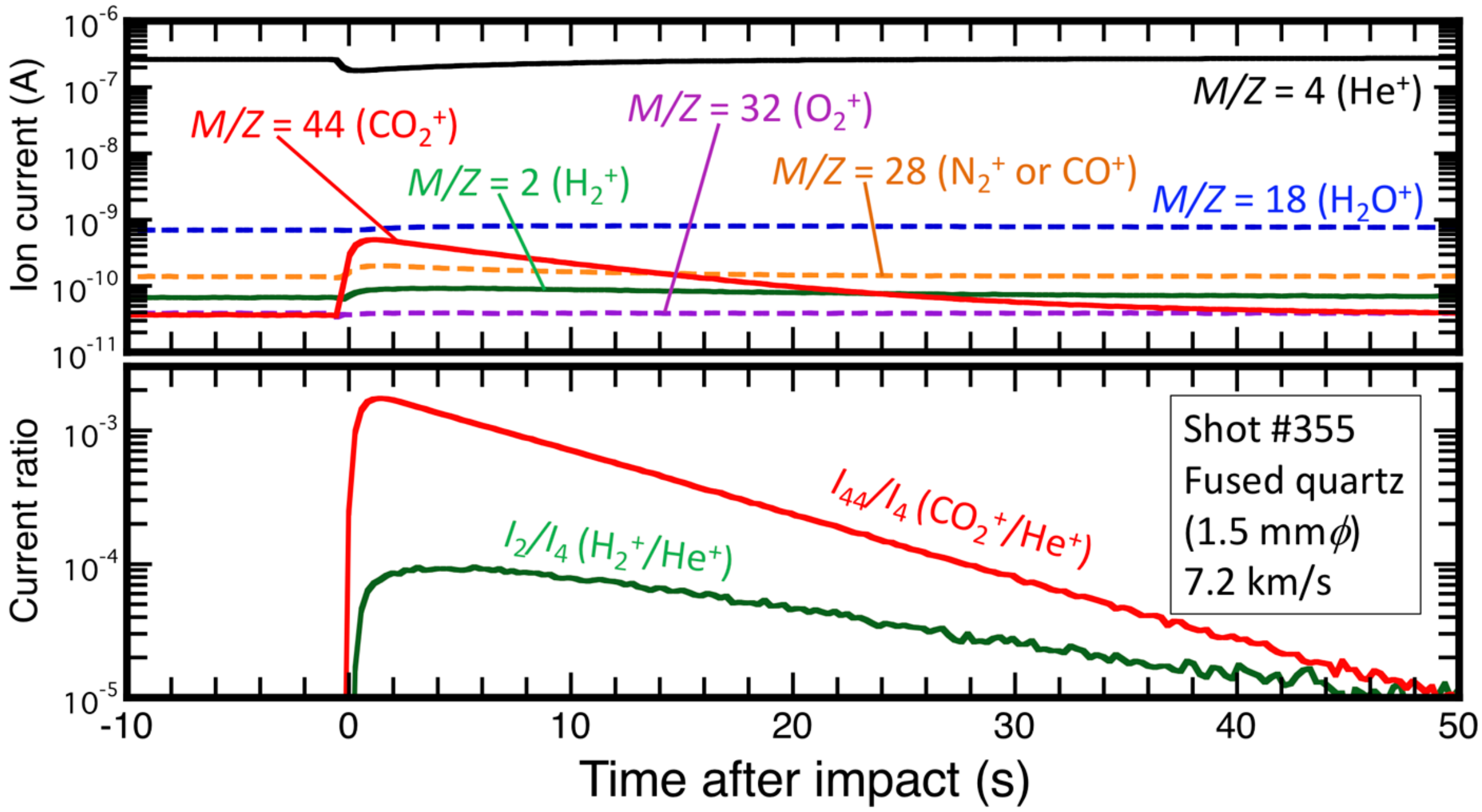

**Figure S6.** An example of the measurement by the quadrupole mass spectrometer in the calcite experiment. (a) The time variation of the ion currents of the selected gas species. (b) The ion current ratio with respect to the ion current for $He^+$. The impact conditions are shown in the figure.

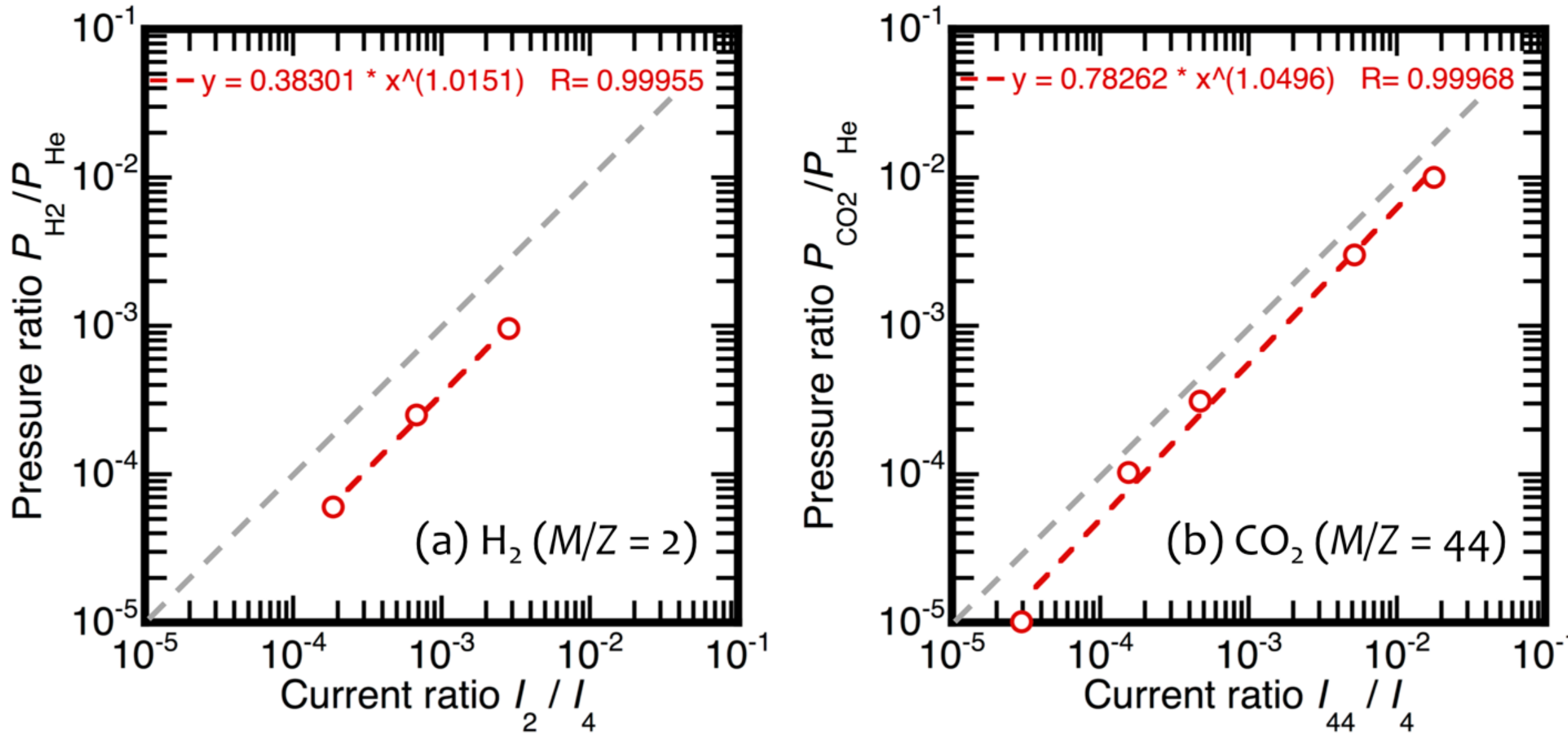

**Figure S7.** Ratio of the partial pressure of (a) $H_2$ ($M/Z$ = 2) and (b) $CO_2$ ($M/Z$ = 44) to He ($M/Z$ = 4). The data points are indicated as open red circles. The red dashed line on each panel is the best-fit line determined by the least-square method. The fitting formula are shown in the figures. The grey dashed line on each panel are the line in the case, which the pressure ratio coincides the ion current ratio.

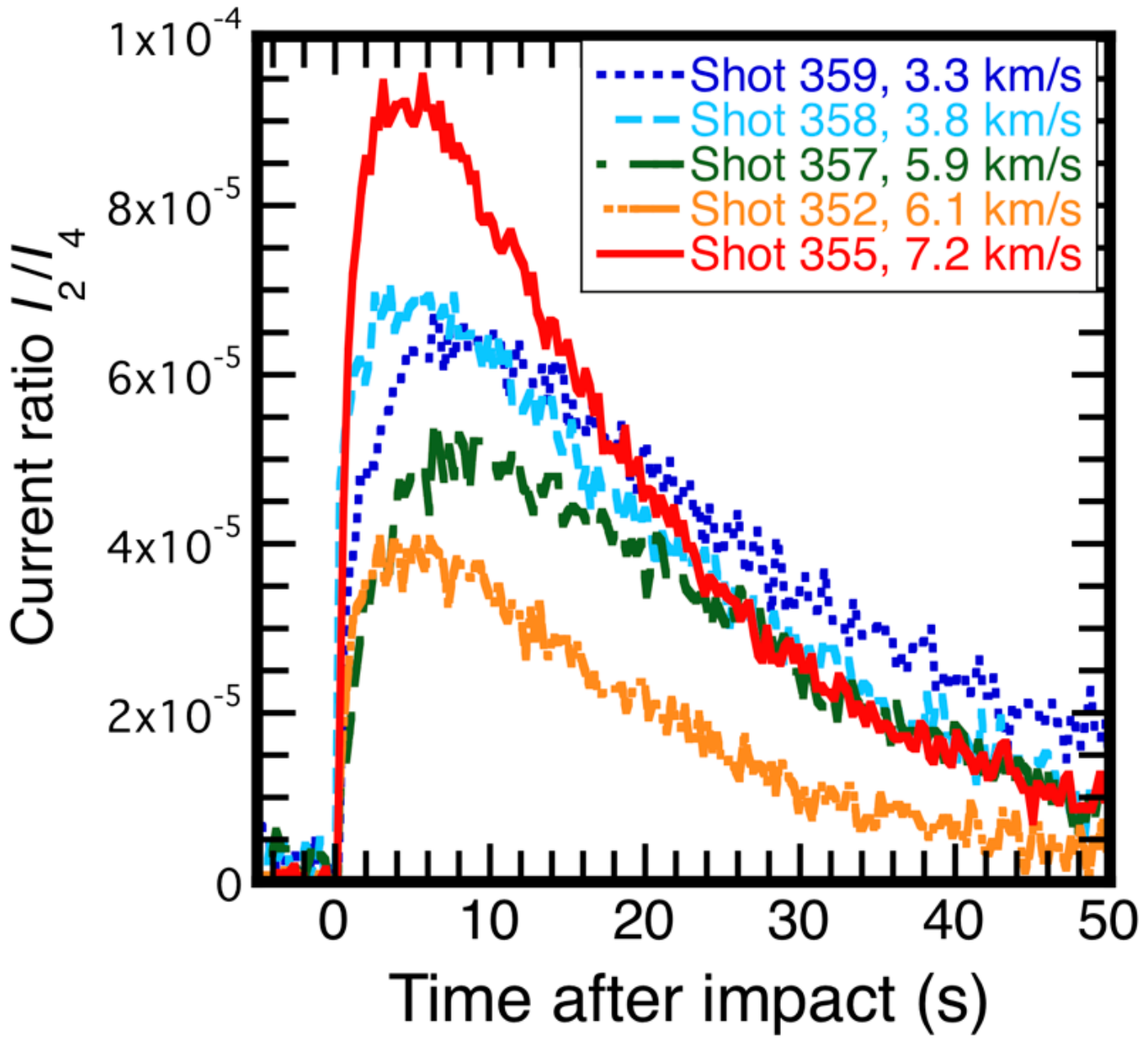

**Figure S8.** The degree of the chemical contamination from the gun. Time variations of the ion current ratios of $H_2^+$ to He+ are shown. The measured projectile speeds for the lines are shown in the figure. The fused quartz projectile with diameters of 1.5 mm (shot #352 and #355) and 2 mm (shot #358 and #359) were used. For the shot #357, we used a $Al_2O_3$ projectile with a diameter of 1 mm.

*S3.5 Gas escape from the experimental chamber*

In the experiments with the two-valve method, the peak ion current for the species of interest, such as $H_2O$ in the experiments on gypsum targets, did not necessarily positively correlate with the impact velocity. This is possibly due to gas escape from the experimental chamber, because valve #2 is opened immediately after the impact. Thus, there is a hole that is ~30 mm in diameter at the entrance of the experimental chamber prior to closing valve #2. In order to examine the degree of gas escape, we conducted two impact experiments with and without the two-valve method. In the control experiment, we used the previous experimental method of Kurosawa et al. (2012) using a plastic diaphragm. We used a projectile that was 1 mm in diameter and smaller than for the experiments described in the main text. In the experiment with the diaphragm, the diameter of the hole pertaining to the gas escape is greatly reduced from 30 to 1 mm, because the hole is produced by penetration of the diaphragm by the projectile. We confirmed that the size of the hole in the plastic diaphragm was nearly identical to the projectile diameter. We used an $Al_2O_3$ projectile to minimize the degree of projectile damage. Figure S9 shows the temporal variations of $I_{44}/I_4$ in the two shots, with and without the plastic diaphragm, at similar impact velocities of ~5.8 km/s. The difference between the two shots is obvious. The peak value of $I_{44}/I_4$ with the diaphragm is four times greater than for the shot without the diaphragm, implying that ~75% of the generated $CO_2$ escaped from the chamber prior to closing valve #2. Consequently, accurate measurement of the total gas production using the two-valve method is still challenging. As such, we only discussed the initiation of shock vaporization/devolatilization with increasing 1-D pressure in the main text.

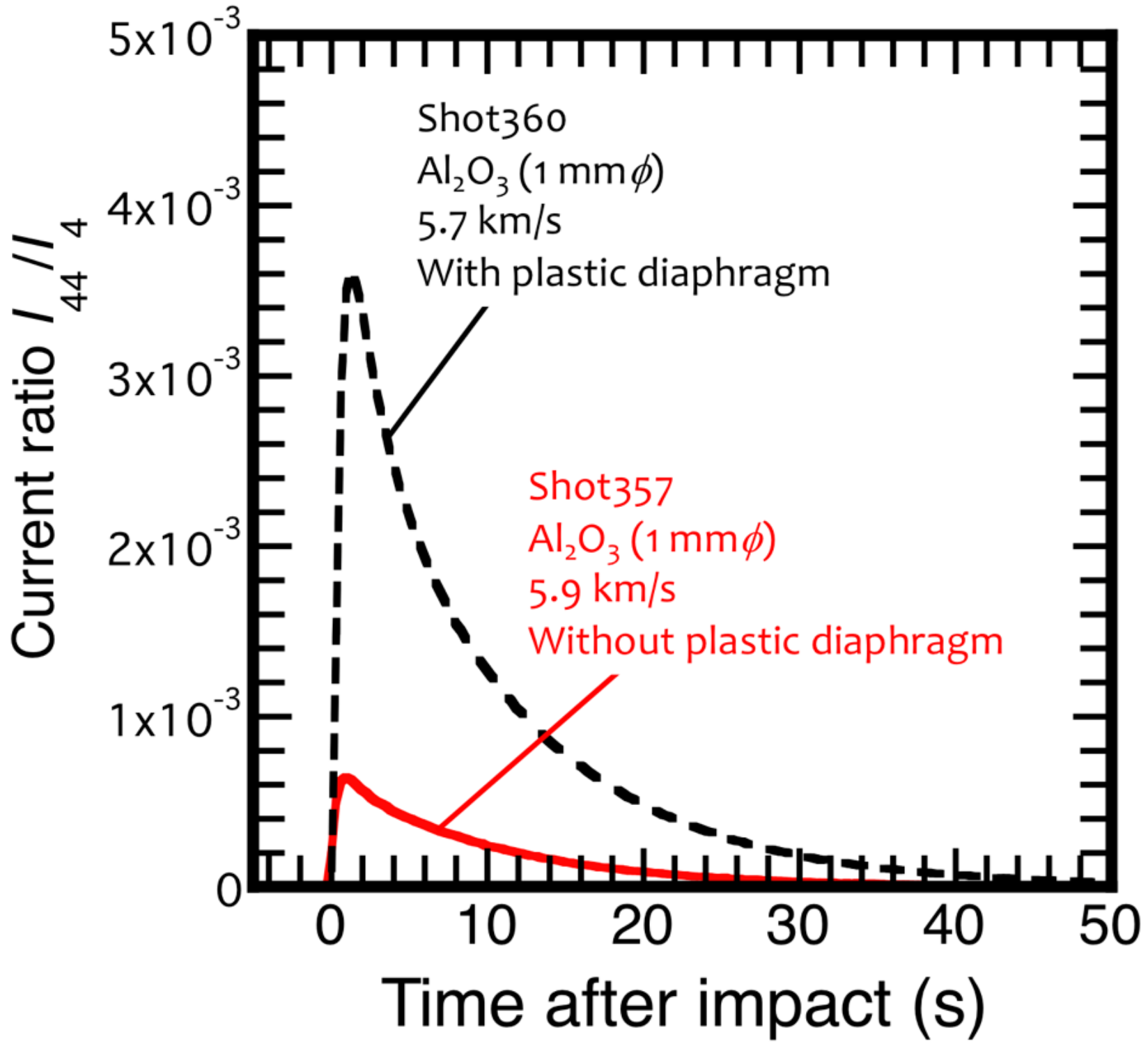

**Figure S9.** The degree of the gas escape. Time variations of the ion current ratios of $CO_2^+$ to He+ are shown. The impact conditions for two shots are indicated in the figure. The red solid line shows the result with the two-valves method. The black dashed line shows the result by using the previous method proposed by Kurosawa et al. (2012).

*S3.6 Detection of shock-generated $CO_2$ at a peak pressure of 6 GPa*

Figure S10 shows the temporal variation in $I_{44}/I_4$ at the lowest impact velocity (1.1 km/s) performed in this study. We clearly detected an increase in $I_{44}/I_4$, although the peak value of $I_{44}/I_4$ was only $2 \times 10^{-5}$. We also conducted a control experiment using a natural basalt block at a slightly higher velocity of 1.4 km/s to rule out the possibility that the generated $CO_2$ in the calcite experiment comes from other sources, such as adsorbed $CO_2$ on the chamber wall and/or floor. No rise in $I_{44}/I_4$ was detected in this control experiment. Thus, the generated $CO_2$ was produced by the shock devolatilization of calcite, even at velocities of 1.1 km/s. The calculated 1-D pressure is only 6 GPa, which is much smaller than the 1-D pressure required for the incipient devolatilization $P_{id}$ estimated by the ANalytical Equation of State (ANEOS) (~50 GPa) (Pierazzo et al., 1998) and about one-third of $P_{id}$ estimated in previous experimental studies (~20 GPa) (e.g., Boslough et al., 1982; Ohno et al., 2008; Kurosawa et al., 2012). Kurosawa et al. (2012) suggested that $P_{id}$ estimated by ANEOS indicates "the effective decarbonation pressure of calcite" (i.e., the pressure required for massive decarbonation by equilibrium shock heating), and that the decarbonation at a lower 1-D pressure is caused by localized energy concentration. Our new results (i.e., detection of decarbonation at ~6 GPa) further highlight the importance of localized energy concentration due to jetting (e.g., Kieffer, 1977) and/or shear banding (Kondo and Ahrens, 1983) and of the heating due to plastic deformation (Kurosawa and Genda, 2018).

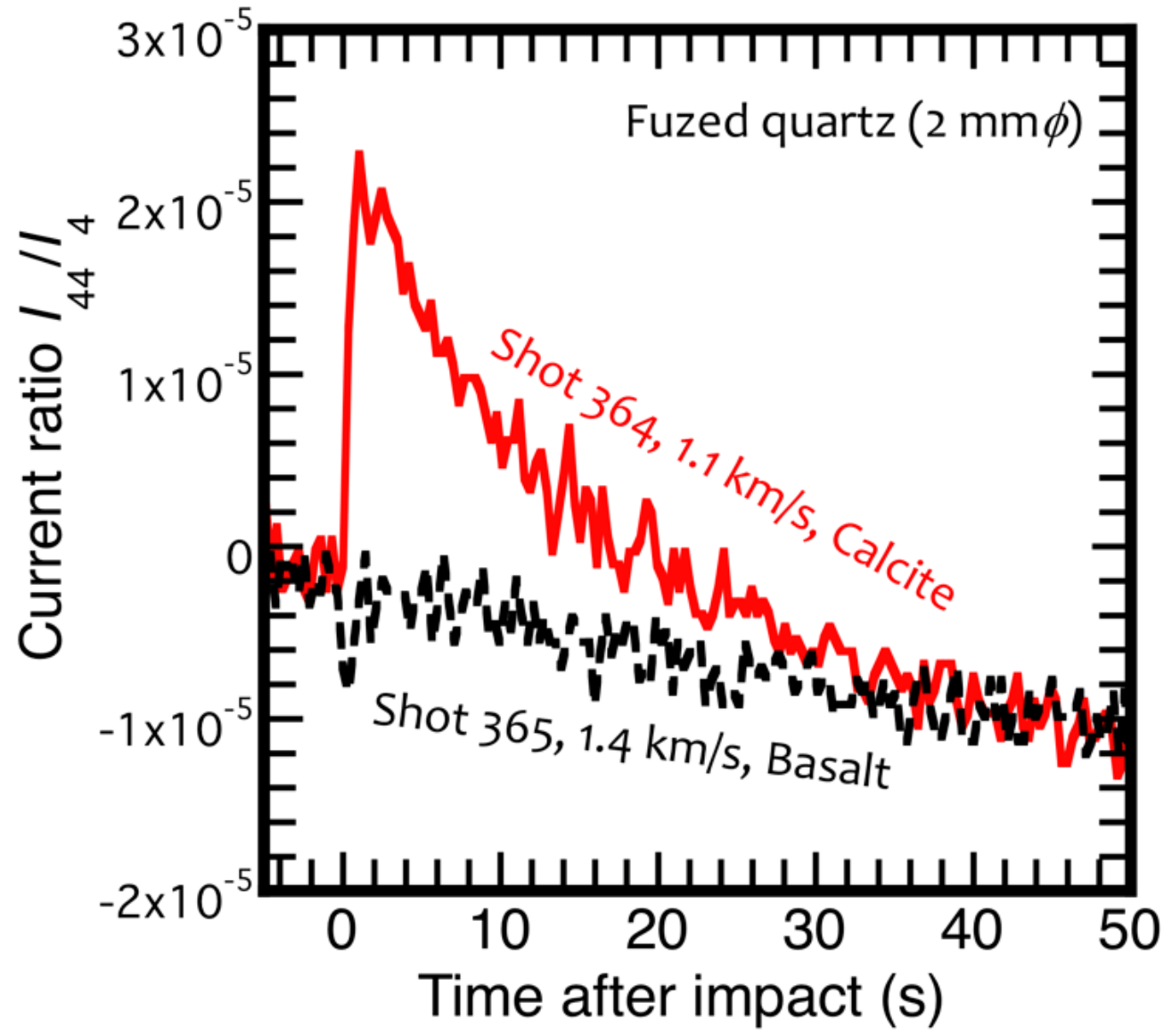

**Figure S10.** Same as Figure S6 except that the result at the lowest impact velocity performed in this study was shown here (red line). The black line shows the result from the control experiment with a basalt block (See Text S3.6)

**Text S4. NaCl thermochemical equilibrium**

We conducted a thermochemical equilibrium calculation to estimate the boiling point of NaCl under the experimental conditions (500 Pa), and the possible chemical composition of the generated vapor. The Lewis code (Gordon & McBride, 1994) involving the Gibbs energy minimization method to calculate the equilibrium chemical composition was used for these calculations. We also included a trace amount of He gas to stabilize the calculation at low temperatures. The atomic composition was set to Na:Cl:He = 1:1:1.1 × $10^{-5}$. Figure S11 shows the molar fraction as a function of temperature. The boiling point of pure NaCl at 500 Pa was estimated to be 1,200 K. Thus, we compared the calculated post-shock temperature to 1,200 K in Fig. 4 of the main text. The thermochemical calculation also provides information about the chemical composition of the vapor phase, showing that the major gas produced is NaCl vapor and that the next most abundant species is $Na_2Cl_2$. As shown in Fig. 2a in the main text, we detected $Cl_2^+$ in the halite experiments. However, the $Cl_2^+$ possibly originated from cracking of $Na_2Cl_2$ by electron impacts in the QMS, and not from $Cl_2$ vapor.

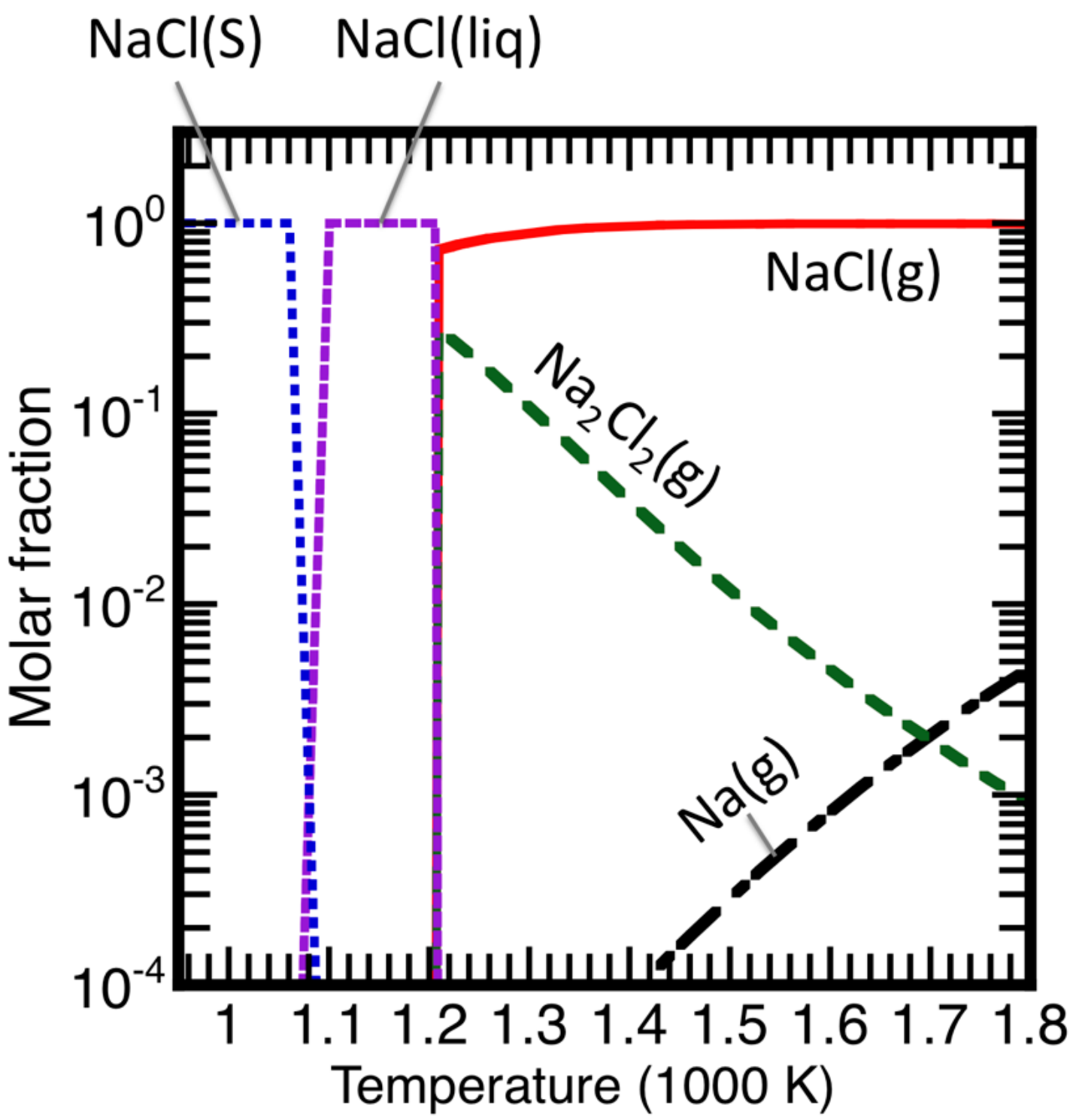

**Figure S11.** Molar fraction of the species with the atomic composition of Na : Cl : He = 1 : 1 : 1.1 x$10^{-5}$ as a function of temperature at 500 Pa. The names of the species are indicated beside the lines in the figure.